\documentstyle[12pt]{article}
\newcommand{\bd}{\begin{document}}
\newcommand{\ed}{\end{document}}
\newcommand{\bc}{\begin{center}}
\newcommand{\ec}{\end{center}}
\newcommand{\be}{\begin{eqnarray}}
\newcommand{\ee}{\end{eqnarray}}
\renewcommand{\thefootnote}{\alph{footnote}}
\newcommand{\se}{\section}
\newcommand{\sse}{\subsection}
\newcommand{\bi}{\bibitem}
\def\figcap{\section*{Figure Captions\markboth
     {FIGURECAPTIONS}{FIGURECAPTIONS}}\list
     {Figure \arabic{enumi}:\hfill}{\settowidth\labelwidth{Figure 999:}
     \leftmargin\labelwidth
    \advance\leftmargin\labelsep\usecounter{enumi}}}
\let\endfigcap\endlist \relax

\begin{document}

\begin{titlepage}
 \vskip 0.5in
 \null
\begin{center}
 \vspace{.15in}
{\Large {\bf Rare kaon decays in SUSY with non-universal $A$ terms
}}\\
\vspace{1.0cm}  \par
 \vskip 2.1em
 {\large
  \begin{tabular}[t]{c}
{\bf Chuan-Hung Chen}
\\
\\
   {\sl Institute of Physics, Academia Sinica, Taipei,}
\\   {\sl  $\ $Taiwan 115, Republic of China }
\\
   \end{tabular}}

 \par \vskip 5.3em
\date{\today}
 {\Large\bf Abstract}
\end{center}
 We study the rare kaon decays in the framework of general SUSY
models. Unlike the results in the literature, we find the
contributions from the gluino exchange to the branching ratio of
$K^+\to \pi^+ \nu \bar{\nu}$ can reach the central value ($\sim
1.5 \times 10^{-10}$) of the new E787 data while the predicted
value of standard model is less than $10^{-10}$. We also find that
the same effects also enhance the decays of $K_{L}\to \pi^0 \nu
\bar{\nu}$, $K_L\to\pi^0e^{+} e^{-}$ and $K_L\to\mu^+ \mu^-$.

\end{titlepage}

One of the essential reasons for the success of the standard model
(SM) is that it naturally satisfies all measured phenomena of
flavor changing neutral current (FCNC) in $K$ and $B$ meson
systems. The FCNC processes are forbidden at the tree level and
suppressed at the loop level by the Glashow-Iliopoulous-Maiani
(GIM) mechanism and by the small quark mixing matrix elements
which involve the transitions between the third and the first two
generations in the Cabibbo-Kobayashi-Maskawa (CKM ) matrix
\cite{CKM}. Therefore, those models which have no GIM mechanism
and flavor mixing suppression could largely enhance the FCNC
decays. Through these decays, one can search for the existence of
physics beyond the SM.

Recently, the processes associated with $|\Delta B|=1$ and
$|\Delta K|=1$ have further progresses in experiments. The
representative of the former is the decay modes of $B \to K \ell^+
\ell^-\ (\ell=e,\mu)$, observed \cite{Belle} at the Belle detector
in the KEKB $e^+e^-$ storage ring with the branching ratio (BR) of
$Br(B\to K \ell^+ \ell^-)=(0.75^{+0.25}_{-0.21}\pm0.09)\times
10^{-6}$, while the SM expectation is around $0.5\times 10^{-6}$
\cite{Bdecays}. The latter is  $K^+\to \pi^+ \nu \bar{\nu}$. The
new data on BR
 from E787 \cite{E787} at the Alternating Gradient Synchrotron
(AGS) of Brookhaven National Laboratory is given by
\begin{eqnarray}
Br(K^+\to \pi^+ \nu \bar{\nu})=(1.57^{+1.75}_{-0.82})\times
10^{-10},
\end{eqnarray}
but the predicted BR in the SM is $(0.72\pm 0.21)\times 10^{-10}$
\cite{DI,Buras}.

By comparing the theoretical uncertainties in both processes, one
can find that the BRs of $B\to K \ell^+ \ell^-$ involve three
independent form factors and their errors can not be reduced by
the exiting measurements. Therefore, before studying the new
physics effects on such processes, it is necessary to know how
large the true theoretical uncertainties are. On the contrary,
$K^+\to \pi^+\nu \bar{\nu}$ has less errors from the hadronic
matrix element and is free of the long-distance uncertainty
\cite{LD}. For eliminating the effect of $K^+\to \pi^+$ matrix
element, the BR of $K^+\to \pi^+ \nu \bar{\nu}$ can be related to
that of $K^+\to \pi^0 e^+ \nu$ with measured BR of 0.0482 by using
the isospin symmetry. Even the corrections to the isospin limit
have also been included by Ref. \cite{MP}. Besides $K^+\to \pi^+
\nu \bar{\nu}$, the relevant decays, such as $K_L\to \mu^+ \mu^-$,
$K_{L}\rightarrow \pi^0 \nu \bar{\nu}$, and $K_{L}\rightarrow
\pi^{0} e^+ e^-$, have the similar characters; especially the BRs
of the last two are related to CP violation (CPV).  By means of
loop induced effects and less hadronic uncertainties, such kind of
rare kaon decays provides good candidates to  test the SM
\cite{BG}. On the other hand, it was pointed out by Ref.
\cite{BB-PLB} that $\beta$ or $\phi_{1}$, one of the three angles
in the CKM matrix, can be described by combining the BRs of
$K^+\to \pi^+ \nu \bar{\nu}$ and $K_{L}\to \pi^0 \nu \bar{\nu}$,
denoted by $[\sin2\phi_{1}]_{\pi \nu \bar{\nu}}$. As well known,
this angle is also related to the time-dependent CP asymmetry of
the $B\to J/\Psi K_{s}$ decay, expressed by
$[\sin2\phi_{1}]_{J/\Psi K_{s}}$. It is obvious
$[\sin2\phi_{1}]_{\pi \nu \bar{\nu}}=[\sin2\phi_{1}]_{J/\Psi
K_{s}}$ in the SM. However, once introducing new CP phases, the
identity will be changed so that
$[\sin2(\phi_{1}+\theta_{K})]_{\pi \nu \bar{\nu}}\neq
[\sin2(\phi_{1}+\theta_{B})]_{J/\Psi K_{s}}$ where $\theta_{K}$
and $\theta_{B}$ are the effects of new physics on  K  and $B$
decays, respectively. Hence, by comparing  $\sin2\phi_{1}$
measured from rare kaon decays and  the asymmetry in the $B$
meson system, we can also tell whether there exist new physics.

Supersymmetric (SUSY) theory not only supplies an elegant
mechanism for the breaking of the electroweak symmetry and a
solution to the hierarchy problem, but possesses abundant flavor
and CP structure. Besides the original CKM matrix, the SUSY models
introduce new flavor mixing effects, such as the upper and down
type squark mixing matrices. The new CP violating phases in SUSY
models can arise from the trilinear and bilinear SUSY soft
breaking $A$ and $B$ terms, the $\mu $ parameter for the scalar
mixing as well as gaugino masses. Unfortunately, it has been
shown that with the universal assumption on the soft breaking
parameters, these phases are severely bounded by electric dipole
moments (EDMs) \cite{Garistosusy} so that the contributions to
$\epsilon $ and $\epsilon ^{\prime }$ are far below the
experimental values. In the literature, some strategies to escape
the constraints of EDMs have been suggested. They are mainly (a)
by setting the squark masses of the first two generations to be as
heavy as few TeV \cite {BKMW} but allowing the third one to be
light; (b) by including all possible contributions to EDMs such
that somewhat cancellations occur in some allowed parameter space
\cite{IN,BGK}; and (c) with the non-universal soft $A$ terms
instead of universal ones. In particular, those models with
non-universal parameters have been demonstrated that they can be
realized in some string-inspired models
\cite{String1,String2,String3,String4}. Moreover, without the
universal assumption, the corresponding off-diagonal terms for the
left-right mixing of the squark mass matrix are unnecessary to be
proportional to the light quark mass directly \cite {String2}. In
this paper, we will show the implication of the generalized $A$
terms on the BRs of $K^+\to \pi^+ \nu \bar{\nu}$ and other
relevant rare kaon decays.

For simplicity, we adopt the mass-insertion approximation in which
the masses of squark are taken as degenerate approximately
\cite{Hall}. We will only concentrate on the $Z^{\mu}\bar{d}
\gamma_{\mu}P_{L(R)}s$ effective interactions introduced by
Z-penguin diagrams with $P_{L(R)}=(1\mp \gamma_5)/2$. Although box
diagrams also contribute to $s\to d \nu \bar{\nu}$, one can easily
check that the effects compared to those of the SM are suppressed
by $M^2_W/M^2_X$ with $M_X$ being the typical SUSY mass scale.
Here, we will not discuss them. As to the dipole vertices such as
$Z^{\mu}\bar{d}\sigma_{\mu\nu}q^{\nu}s$, because they are
suppressed by $q/M_Z$ and by the light quark masses $m_{s(d)}$,
their contributions are also negligible. Hence, there are two
mechanisms to generate effective couplings for $Z^{\mu}\bar{d}
\gamma_{\mu}P_{L(R)}s$. One is from the penguin of chargino
exchange \cite{BM,BRS,CI,BCIRS} and another one is from the same
diagram but with gluino exchange \cite{BM}. Comparing to the
contributions of chargino and gluino, due to smaller couplings,
the effects of neutralino are always negligible.

In the literature it is claimed that due to the magnitude of the
relevant mass-insertion parameters being proportional to the light
quark masses \cite{BRS,CI,BCIRS} or proportional to the suppressed
quark mixing matrix elements \cite{BM}, the effects from gluino
are much smaller than those of the SM. As a consequence, the
dominant one comes from the chargino sector that the
mass-insertion terms are associated with stop-quark effects and
the corresponding constraints on the squark mixing matrix elements
are looser \cite{CI}. However, it is shown by the new data of E787
that the central value has been close to the predicted value of
the SM. In spite of the still large errors, if the central value
hints the existence of new effects, actually it implies that the
models with very large contributions to $K^+\to \pi^+\nu
\bar{\nu}$ should be further constrained. On the contrary, it
increases the possibility for the new physics which is compatible
with the SM. Due to the enormous progress in the SUSY flavor
physics, what we want to emphasize is that the contributions from
the Z-penguin with gluino exchange are not negligible and can have
sizable effects if the conventional soft SUSY breaking $A$ terms
possess more general flavor structures or are non-universal
(nondegenerate). To accomplish the purpose, in the following
analysis, we consider the case that the contributions from
chargino are small.

We start by writing the effective interactions for $Z$ coupling to
quarks and squarks as
\begin{eqnarray}
{\cal L}&=&-i{g\over c_{W}}\sum_{f}\Big[ \tilde{f}^*_{L}
\stackrel{\leftrightarrow}{\partial}_{\mu}
(T_{3f}-s^2_{W}Q_{f})\tilde{f}_{L} \nonumber \\ &&-
\tilde{f}^*_{R}
\stackrel{\leftrightarrow}{\partial}_{\mu}(s^2_{W}Q_{f})\tilde{f}_{R}
\Big]Z^{\mu} - {g\over c_{W}}\sum_{f}\bar{f}\gamma_{\mu} \nonumber
\\ && \Big[ (T_{3f}-s^2_{W}Q_{f})P_{L}- s^2_{W}Q_{f} P_{R} \Big] f
Z^{\mu},\label{zff}
\end{eqnarray}
where $f$ denotes the fermion and $\tilde{f}_{L(R)}$ corresponds to
the superpatner of $f$ with the chirality L(R),
$s_{W}(c_{W})=\sin\theta_{W}(\cos\theta_{W})$ with $\theta_{W}$
being the Weinberg angle and $Q_{f}$ is the charge of $f$. On the
other hand, the relevant squared down type squark mass matrix is
described by
\begin{eqnarray}
{\cal M}^{2}_{\tilde{D}}=\left( \begin{array}{cc}
  ({\bf m}^2_{\tilde{D}})_{LL} & ({\bf m}^2_{\tilde{D}})_{LR}\\
 ({\bf m}^2_{\tilde{D}})^{\dagger}_{LR} & ({\bf m}^{2}_{\tilde{D}})_{RR}
     \\
 \end{array}\right),
\end{eqnarray}
\begin{eqnarray}
({\bf m}^2_{\tilde{D}})_{LL}&=&({\bf M}^2_{\tilde{D}})_{LL}+{\bf
m}^2_{D}-{\cos2\beta\over 6}(M^2_{Z}+2M^2_{W}){\bf \hat{1}},
\nonumber \\ ({\bf m}^2_{\tilde{D}})_{LR}&=&({\bf
M}^2_{\tilde{D}})_{LR}-\mu^* \tan\beta {\bf m}_{D}, \nonumber \\
({\bf m}^2_{\tilde{D}})_{RR}&=&({\bf M}^2_{\tilde{D}})_{RR}+{\bf
m}^2_{D}-{\cos2\beta\over 3}M^2_{Z}\sin^{2}\theta_{W}{\bf
\hat{1}}\,,
\end{eqnarray}
where we have adopted the so-called super-CKM basis that the
quarks have been the mass eigenstates so that  ${\bf m}_{D}$ is
the diagonalized down quark mass matrix, ${\bf \hat{1}}$ denotes
the $3\times 3$ unit matrix, the definition of the angle $\beta$
is followed by $\tan\beta=v_{u}/v_{d}$ with $v_{u}$ and $v_{d}$
being the vacuum expectation values (VEVs) of Higgs fields
$\Phi^u$ and $\Phi^d$ responsible for the masses of upper and down
type quarks, respectively, $\mu$ is the mixing effect of $\Phi^u$
and $\Phi^d$, $({\bf M}^2_{\tilde{D}})_{LL(RR)}$ stand for the
soft breaking masses for down type squarks and $({\bf
M}^2_{\tilde{D}})_{LR}$ describe the trilinear soft breaking
couplings and are written as
\begin{eqnarray}
({\bf M}^2_{\tilde{D}})_{LR}={v_{d}\over \sqrt{2}}
V_{D_{L}}\tilde{A}^{d*} V^{\dagger}_{D_{R}}, \label{lrm}
\end{eqnarray}
where $V_{D_{L(R)}}$ transform the left(right)-handed quarks from
weak eigenstates to mass eigenstates and
$\tilde{A}^{d}_{ij}=Y^{d}_{ij}A^{d}_{ij}$ with $Y^{d}_{ij}$ and
$A^{d}_{ij}$ being Yukawa and soft SUSY breaking matrix,
respectively.

Compared to the $M_{W}$ scale, due to the smallness of the
involved momentum or momentum transfer and  the masses of external
legs, we drop them in our considerations so that one can easily
get the conclusion by the similar situation to the gauge invariant
requirement on $\gamma-s-d$ vertex that the one-loop contributions
 from the left-left (LL) mixing of the down squark mass matrix are
vanished. Therefore, in order to get the effective interaction
$Z-s-d$ that the incoming and outgoing particles carry the same
chirality, it needs a double mass-insertion in the squark
propagator, {\it i.e.}, the appearance of
$\sum_{j}(M^2_{\tilde{D}2j})_{AB}(M^2_{\tilde{D}j1})_{BA}$ with
$A(B)=L(R)$ or $R(L)$ and  $j=1,2,3$ is necessary \cite{BM}.
In  general, we also consider $\tilde{g}-\tilde{q}_{L}-q_{L}$ and
$\tilde{g}-\tilde{q}_{R}-q_{R}$ vertices simultaneously. The
situation is different from the case of chargino exchange in which
only the left-handed couplings give the main contributions.

According to the interactions in Eqs. (\ref{zff}) and (\ref{lrm}),
by including the self-energy diagrams and all possible emissions
of $Z$-boson from the propagators of internal squarks, the
effective interactions for $s\rightarrow d Z$ can be derived as
\begin{eqnarray}
{\cal L}_{\tilde{g}}&=& C_{Z} [ \tilde{Z}_{L}(x)\,
\bar{d}\gamma_{\mu} P_L s \, - \tilde{Z}_{R}(x)\,
\bar{d}\gamma_{\mu} P_R s ]  Z^{\mu}\label{effz}
\end{eqnarray}
with
\begin{eqnarray}
C_{Z}&=&{G_F e\over \sqrt{2} \pi^2} M^{2}_{Z}{\cos\theta_{W} \over
\sin\theta_{W}}, \nonumber \\
 \tilde{Z}_{A}(x)&=& { \alpha_{s}\sin^{2}\theta_{W}
\over 12 \alpha_{em}}C_F G(x) \sum_{j=1,2,3}(
\delta^{d}_{2j})_{AB} ( \delta^{d}_{j1})_{BA},\label{effzp}
\end{eqnarray}
where $C_F=4/3$, $(\delta _{ij}^{d})_{AB}\equiv
(M^2_{\tilde{D}ij})_{AB}/m_{\tilde{q }}^{2}$, $m_{\tilde{q}}$ is
the average mass of squark in the super-CKM basis and
\begin{eqnarray*}
 G(x)&=&{2x^2+5x-1 \over 2(x-1)^{3}} -{3x^2\ln x \over
(x-1)^{4}}
\end{eqnarray*}
with $x=m^2_{\tilde{g}}/m^2_{\tilde{q}}$ and $m_{\tilde{g}}$ being
the gluino mass. We note that according to  Eq. (\ref{effz}) the
effect of the different chirality is opposite in sign each other.
As known, the associated matrix elements for relevant kaon decays
is $<\pi|\bar{d}\gamma_{\mu}s|K>$ so that the opposite sign
actually reflects somewhat cancellation between different chiral
couplings. We will see later that in some SUSY models, the
cancellation is almost complete.

Altogether, the effective interactions combined with those of the
SM can be written as
\begin{eqnarray}
{\cal L}&=& C_{Z} \Big[ \Big(X_{SM}(x_t)+\tilde{Z}_{L}(x)\Big)\,
\bar{d}\gamma_{\mu} P_L s \nonumber \\ &&
 - \tilde{Z}_{R}(x)\,
\bar{d}\gamma_{\mu} P_R s \Big ] Z^{\mu}
\end{eqnarray}
with
\begin{eqnarray*}
 X_{SM}(x_t)&=&\lambda_{c}P_0+\lambda_{t}X_{0}(x_t).
\end{eqnarray*}
The explicit expressions of functions $X_{0}(x_t)$ and
$Y_{0}(x_t)$ can be found in Ref. \cite{BCIRS}. $P_0$ is given in
Ref. \cite{BB}. From the formulas in Ref. \cite{BCIRS}, the BR
is expressed as


\begin{eqnarray}
Br(K^+\to\pi^+\nu_l\bar{\nu}_l)&=& {1\over 3}\kappa_+ \Big[
(ImF)^{2}+(ReF)^{2} \Big]\label{rarek}
\end{eqnarray}
with
\begin{eqnarray}
F&=&{\lambda_c\over\lambda}P_0+{\lambda_t\over\lambda^5}X_{0}(x_t)
+{1 \over\lambda^5}X_{\tilde{g}}(x),
\nonumber \\%
\kappa_+&=&r_{K^+}{3\alpha^2 Br(K^+\to\pi^0e^+\nu)\over 2\pi^2
\sin^4\theta_W}
 \lambda^8,\nonumber \\
X_{\tilde{g}}(x)&=&\tilde{Z}_{L}(x)-\tilde{Z}_{R}(x),
\label{zlr}
\end{eqnarray}
where $x_{t}=m^{2}_t /m^{2}_{W}$, $\lambda_i=V^\ast_{is}V_{id}$
with $\lambda_c$ being real to a very high accuracy, and
$r_{K^+}=0.901$ summarizes isospin breaking corrections in
relating $K^+\to\pi^+\nu\bar{\nu}$ to $K^+\to\pi^0e^+\nu$. Here,
the isospin breaking corrections from the quark mass effects and
electroweak radiative corrections have been calculated in Ref.
\cite{MP}. By means of Eq. (\ref{zlr}), we clearly see that
if the $SU(2)_{L}$
breaking effects for left- and right-handed coupling are the same,
the influence on  $Br(K^+ \to \pi^+ \nu \bar{\nu})$  also
vanishes. With the same effective interactions, the contributions
 from the gluino exchange to BRs of  $K_{L}\to \pi^0 e^+ e^-$,
$K_{L}\to \pi \nu \bar{\nu}$ and  $K_{L}\to \mu^+ \mu^-$can also
be described by

\begin{eqnarray}
 Br(K_L\to\pi^0e^{+} e^{-})_{dir}&=& \kappa \Big[ |Imy_{7A}|^{2} +
|Imy_{7V}|^{2} \Big]\nonumber \\
Br(K_L\to\pi^0\nu_l\bar{\nu}_l)&=& {1\over 3}\kappa_L
(ImF)^{2} \nonumber \\ %
Br(K_L\to\mu^+ \mu^-)_{SD}&=& \kappa_{\mu} (ReD)^{2}
\label{brrarek}
\end{eqnarray} %
 with
\begin{eqnarray*}
y_{7A}&=& \frac{\lambda_{t}}{\lambda^5}  Y_{0}(x_t) + \frac{1}{\lambda^5} X_{\tilde{g}}(x), \nonumber \\
y_{7V}&=& \frac{\lambda_{t}}{\lambda^5}
[1+(1-4\sin^2\theta_{W})C_{0}(x_t)] \nonumber \\
&&+\frac{1}{\lambda^5}
(1-4\sin^2\theta_{W})X_{\tilde{g}}(x),\nonumber \\
D&=&{\lambda_{t}\over \lambda^5}Y_0(x_t)+{\bar{\Delta}_{c}\over
\lambda^5}+{X_{\tilde{g}}\over \lambda^5}
\end{eqnarray*}
and
\begin{eqnarray*}
\kappa_L&=&r_{K_L}{\tau_{K_L}\over \tau_{K^+}}{3\alpha^2
Br(K^+\to\pi^0e^+\nu)\over 2\pi^2 \sin^4\theta_W}
 \lambda^8=1.89\cdot 10^{-10}\, ,
\end{eqnarray*}
where $\kappa =\kappa_{L}/6$, $\kappa_{\mu}=1.68\times 10^{-9}$,
and $C_{0}(x_t)$ and $\bar{\Delta}_{c}$ from the charmed loop can
be found in Ref. \cite{BCIRS}.

 The essential question is whether the involving gluino
contributions can yield the BR of $K^+\to \pi^+ \nu \bar{\nu}$ as
large as that given by  E787. To explore the possibility, we
have to analyze the constraints on the relevant parameters. For
simplicity, we just consider the case for j=3 in Eq. (\ref{zlr})
so that the involving mass-insertion parameters are only
$(\delta^{d}_{23})_{AB}$ and $(\delta^{d}_{31})_{AB}$ although
this assumption is unnecessary. To estimate the CP violating
effects,
we take that the real and imaginary parts of relevant parameters
are  approximately the same in order of magnitude, such as
$|Im(\delta^{d}_{ij})_{AB}|\approx |Re(\delta^{d}_{ij})_{AB}|/2$.
Under our assumption, it is known immediately that the bounds on
the parameters are from  $Br(B\to X_{s}\gamma)$ and the
$B_{d}-\bar{B}_{d}$ mixing. The former constrains
$(\delta^{d}_{23})_{AB}$ while the latter is
$(\delta^{d}_{31})_{AB}$. In addition, it is worth mentioning that
charge and color breaking (CCB) minima and the potential unbounded
 from below (UFB) may also give strict bounds \cite {CD}. To relax
the constraints from the vacuum instability, we adopt the
following two strategies: (a) According to the result in Ref.
\cite{CD}, the model independent upper bounds on
$(\delta^{d}_{3j})_{LR}$ can be described by $m_{b}
\sqrt{2+m^{2}_{\tilde{l}}/m^{2}_{\tilde{q}}}/m_{\tilde{q}}$. By
taking the slepton mass to be few TeV, the constraints are
compatible with those directly obtained by FCNC decays. (b) Our
universe is resting a false vacuum. From the analysis in Ref.
\cite{KL}, the constraints from the conditions of CCB minima and
UFB can be relaxed if the life time of the metastable vacuum is as
long as that of the present age of the universe. Although the
requirements of CCB and UFB in \cite{CD} are necessary, after all,
they are not sufficient. Hence, in our numerical calculations, we
use the constraints gotten from FCNC processes.

Since the relevant constraints have been considered in Ref.
\cite{GGMS}, we display the bounds in Table \ref{limit} for fixing
the specific chirality.
\begin{table}[htpb]
\caption{The bounds on the relevant parameters \cite{GGMS}.}
\label{limit}
\begin{center}
\begin{tabular}{cccc}
    \hline
x & $|Re(\delta^{d}_{31})_{RL}|\times 10^{2}$ &
$|Re(\delta^{d}_{23})_{LR}|\times 10^{2}$ & G(x)
    \\ \hline
  0.3 & $4.57 $ & $1.16 $ & 0.36 \\
  1.0 & $3.81 $ & $1.43 $ & 1/4 \\
  4.0 & $4.16$ & $2.68 $ & 0.18 \\
  \hline
\end{tabular}
\end{center}
\end{table}
The values in the entries of Table \ref{limit} are for
$m_{\tilde{q}}=500$ GeV. For the different choices, the values for
the second column need to be multiplied by $m_{\tilde{q}}/500$ but
it is $(m_{\tilde{q}}/500)^{2}$ for the third column.
The parameters that all R(L) are replaced by L(R) depend on
the details of
SUSY models.
 As illustrations, we show
three possible situations in SUSY models:

$\bullet$ Scenario I: $\tilde{Z}_{L} \simeq \tilde{Z}_{R}$. This
approximation is equivalent to $(\delta^{d}_{ij})_{RL}\simeq
(\delta^{d}_{ij})_{RL}$. It has been shown recently that if the
Yukawa and soft-breaking $A$ matrices are hermitian, the identity
could be realized \cite{Mohapatra,ABKL,AKL}. As a result of Eq.
(\ref{zlr}), the contributions to rare kaon decays all vanish.
Nevertheless, this scenario implies that the hyperon CP asymmetry
in SUSY is one order of magnitude larger than that of the SM
\cite{CC}.

$\bullet$ Scenario II: $|\tilde{Z}_{R} - \tilde{Z}_{L}|\sim
|\tilde{Z}_{L(R)}|$. Besides the negligible contributions to rare
K decays in scenario I, another shortcoming is that the predicted
$\epsilon'$ is also far smaller than that of the experimental
measurement \cite{ABKL}. In order to deal with the small
$\epsilon'$ problem and escape the constraint from the EDM, the
asymmetric soft-breaking $A$ matrix is proposed by Ref. \cite{KK}
and it is also found that such kind of the asymmetric property
could be realized in some string-inspired supergravity models.
Hence, the situation of scenario II can be reached if the SUSY
soft-breaking $A$ matrix is asymmetric. According to the results
of Eqs. (\ref{rarek}) and (\ref{brrarek}) and the constraints in
Table \ref{limit}, the BRs of relevant rare kaon decays are found
as follows:
\begin{eqnarray}
Br(K^+\to\pi^+ \nu \bar{\nu})&=&1.55 \times 10^{-10}, \nonumber \\
Br(K_L\to\pi^0\nu\bar{\nu})&=& 2.50 \times 10^{-10}, \nonumber \\
Br(K_L\to\pi^0e^{+} e^{-})_{dir}&=& 3.47 \times 10^{-11},\nonumber \\
Br(K_L\to\mu^+ \mu^-)_{SD}&=& 1.85\times 10^{-9}. \label{br2}
\end{eqnarray} %
Because our purpose is to demonstrate that the gluino
contributions from penguin diagrams to rare kaon decays could
reach the current experimental ranges, the values in Eq.
({\ref{br2}) are obtained by setting $m_{\tilde{q}}=640$ GeV with
$x=0.3$ and the sign of $\tilde{Z}_{L(R)}$ is the same as that of
SM. And also we choose the value of $\lambda_{t}$ such that the
results of the SM for $K^+\to\pi^+ \nu \bar{\nu}$,
$K_L\to\pi^0\nu\bar{\nu}$, $K_L\to\pi^0e^{+} e^{-}$, and
$K_L\to\mu^+ \mu^-$ are $0.58\times 10^{-10}$, $0.46 \times
10^{-10}$, $0.7 \times 10^{-11}$ and $5.26 \times 10^{-10}$,
respectively.

$\bullet$ Scenario III: $\tilde{Z}_{R} \ll \tilde{Z}_{L}$ or vice
versa. In this situation, as expected, the numerical predictions
should be similar to the case of scenario II.

Finally, we give a brief discussion on chargino contributions.
According to the analysis of Ref. \cite{CI}, one can understand
that unlike gluion case, the main effects of chargino on
$Z\bar{d}s$ effective interaction come from the wino component
where
only left-handed couplings are involved. That is, only
$(\delta^{u}_{2j})_{LR}(\delta^{u}_{j1})_{RL}$ have the
significant contributions. However, if the trilinear soft breaking
$A^{u}_{ij}$ and $A^{d}_{ij}$ parameters arise from the same
origin and have the same order of magnitude, due to the smaller
weak couplings, the effects of chargino could be much smaller than
those of gluino if both superparticles have the same masses. On the
other hand, if we allow that $A^{u}$ is different from $A^{d}$
entirely,  in order to guarantee that the effects of chargino are
negligible, the mass of lightest chargino is taken as heavy as
$O(TeV)$ so that although the constraints of relevant mass
insertion parameters are not strict, all contributions from
chargino will become insignificant.

In summary, we have studied the rare kaon decays in the framework
of general SUSY models. In terms of the rich flavor structure, we
find that the effects of the gluino exchange can make $Br(K^{+}\to
\pi^{+} \nu \bar{\nu})$ up to the central value in the new
BNL-E787 result. With the same effects, the BRs of the short
distance contributions to  $K_{L}\to \pi^0 \nu \bar{\nu}$,
$K_L\to\pi^0e^{+} e^{-}$ and $K_L\to\mu^+ \mu^-$ are larger than
those in the SM.\\

\noindent {\bf Acknowledgments}

I would like to thank C.Q. Geng and H.N. Li for their useful
discussions. I also thank G. Isidori for introducing me to the
constraints from CCB minima and potential UFB. This work was
supported in part by the National Science Council of the Republic
of China under Contract No. NSC-90-2112-M-001-069 and the National
Center for Theoretical Science.

\end{document}